\documentclass[useAMS,usenatbib]{mn2e}\usepackage{epsfig}\usepackage{amsmath}

\newcommand{\be}{\begin{equation}}
\newcommand{\beq}{\begin{equation}}
\newcommand{\ba}{\begin{eqnarray}}
\newcommand{\ee}{\end{equation}}
\newcommand{\eeq}{\end{equation}}
\newcommand{\ea}{\end{eqnarray}}

\def\lsim{~\rlap{$<$}{\lower 1.0ex\hbox{$\sim$}}}

\def\gtrsim{~\rlap{$>$}{\lower 1.0ex\hbox{$\sim$}}}

\title[Supernova Feedback in high redshift galaxies]{A suppressed
contribution of low mass galaxies to reionization due to supernova
feedback}

\author[Wyithe \& Loeb]{J. Stuart B. Wyithe$^{1}$, Abraham Loeb$^{2}$\\$^1$
School of Physics, University of Melbourne, Parkville, Victoria,
Australia\\$^2$ Astronomy Department, Harvard University, 60 Garden
Street, Cambridge, MA 02138, USA\\Email: swyithe@unimelb.edu.au; aloeb@cfa.harvard.edu}

\begin{document}


\maketitle

\label{firstpage}
\begin{abstract}

\noindent Motivated by recent observations of the star formation rate
density function out to $z\sim7$, we describe a simple model for the
star formation rate density function at high redshift based on the
extended Press-Schechter formalism. This model postulates a starburst
following each major merger, lasting for a time $t_{\rm SF}$ and
converting at most $f_{\rm \star,max}$ of galactic gas into stars. We
include a simple physical prescription for supernovae feedback that
suppresses star formation in low mass galaxies.  Constraining $t_{\rm
SF}$ and $f_{\rm \star,max}$ to describe the observed star formation
rate density at high redshifts, we find that individual starbursts
were terminated after a time $t_{\rm SF}\sim10^7$ years. This is comparable to
the main-sequence lifetimes of supernova progenitors, indicating that
high redshift starbursts are quenched once supernovae feedback had
time to develop. High redshift galaxies convert $\sim10\%$ of their
mass into stars for galaxies with star formation rates above
$\sim1$M$_\odot~{\rm yr^{-1}}$, but a smaller fraction for lower
luminosity galaxies.  Our best fit model successfully predicts the
observed relation between star formation rate and stellar mass at
$z\gtrsim4$, while our deduced relation between stellar mass and halo
mass is also consistent with data on the dwarf satellites of the Milky
Way. We find that supernovae feedback lowers the efficiency of star
formation in the lowest mass galaxies and makes their contribution to
reionization small. As a result, photo-ionization feedback on low mass
galaxy formation does not significantly affect the reionization
history. Using a semi-analytic model for the reionization history, we
infer that approximately half of the ionizing photons needed to
complete reionization have already been observed in star-forming
galaxies.

\end{abstract}


\section{Introduction}

The galaxy luminosity function is the primary observable that must be
reproduced by any successful model of galaxy formation. At
$z\gtrsim6$, it also represents one of the most important observables
for studying the reionization of cosmic hydrogen.  Developing a
theoretical picture of the important processes involved in setting the
star formation rate at high redshift lies at the forefront of
understanding this important cosmic epoch
\citep[e.g.][]{Trenti2010,Finlator2011,Munoz2011,Raicevic2011,Salvaterra2011}.

The luminosity function of Lyman-break galaxy candidates discovered at
$z\ga6$ in the Hubble Ultra-Deep Field is described by a Schechter
function with characteristic density $\Psi_\star$ in comoving
Mpc$^{-3}$, and a power-law slope $\alpha$ at luminosities $L$ below a
characteristic break $L_\star$
\citep[e.g.][]{Bouwens2011}. Observations show that the value of
$L_\star$ decreases towards higher redshift as expected from the dark
matter halo mass-function \citep[e.g.][]{Munoz2011}, while the faint
end slope of $\alpha\approx -1.8$ is observed to be roughly
independent of redshift \citep[][]{McLure2009,Bouwens2011}.
A complication that arises when modelling the luminosity function is
that models predict a star formation rate, which must then be
converted to a luminosity assuming an initial mass function (IMF) for
the stars. While this calculation is straightforward, of more
importance is the potential contribution of reddening. As shown by
\citet[][]{Bouwens2012}, the blue continuum slope of $z\gtrsim6$
galaxies is dependent on luminosity and redshift. As a result there is
a dust correction which is both luminosity and redshift dependent and
difficult to reproduce from first principles reliably in a simple
model. This renders theoretical studies of the luminosity function
difficult to interpret. For example, in the model of
\citet[][]{Munoz2011} a distribution of mass-to-light ratios is
assumed whose mean is independent of mass, while in
\citet[][]{Trenti2010} the mass-to-light ratio is assumed to be
independent of redshift. Neither of these assumptions holds based on
the more recent observational work of \citet[][]{Bouwens2012}.
More complex models are able to better reproduce many of the observed
properties, and so make more robust predictions of the physics. For
example \citet[][]{Finlator2011} have modelled the growth of stellar
mass in high redshift galaxies using hydrodynamical simulations
coupled with sub-grid models for processes including star formation
and metal enrichment, and broadly reproduce the luminosity function
evolution as well as the blue colours of the young stellar populations
at high redshift.  Similarly, \citet[][]{Salvaterra2011} and
\citet[][]{Jaacks2012} have calculated the evolution of the luminosity
function in detailed numerical simulations including calculations of
enrichment and dust reddening, with the latter also including
additional physics related to the transition from population-III to
population-II stars. A focus of these numerical studies is the role of
supernova (SNe) feedback on star formation, particularly in low mass
systems. In a separate approach, \citet[][]{Raicevic2011} \citep[see
also ][]{Benson2006,Lacey2011} used a semi-analytical galaxy formation
code based on Monte-Carlo merger trees to model the
evolution of the high redshift luminosity function, and study
the effect of SNe feedback on the global ionizing photon budget and
global ionization. In particular \citet[][]{Raicevic2011} evaluated
the ionizing photon budget, finding that although galaxies should
produce sufficient ionizing photons to complete reionization, most of
the galaxies responsible would be below the detection threshold of
current surveys.

A simpler way to constrain theory by observations is to estimate the
star formation rate density observationally, where the correction is
made from luminosity to star formation rate using the observed
continuum properties of the galaxies under study. A powerful probe of
the physics of star formation is then provided by the star formation
rate density function (i.e. the number of galaxies per unit volume per
unit star formation rate).  Recently, this has become a viable
approach following the work of \citet[][]{Smit2012} who combined
estimates of dust extinction at $z\sim 4 - 7$ with measurements of the
UV luminosity function in order to derive star formation rate density
(SFRD) functions at $z \sim4$, 5, 6 and 7. These SFRD functions
provide a physical description of the build-up of stellar mass in
galaxies at high redshift. The resulting SFRD functions are
well-described by a Schechter function, with a characteristic
break separating a shallow dependence of SFRD on star formation rate
at low luminosities from the exponential dependence at high
luminosities.
As mentioned, the physics of star formation in high redshift galaxies
has important implications for the process of reionization. Of
particular interest is the star formation rate in low mass
systems. For example, it has been shown that the growth of HII regions
during reionization may be influenced by radiative feedback in the
form of suppression of galaxy formation below the cosmological Jeans
mass within a heated intergalactic medium (IGM)
\citep[][]{Dijkstra2004b}, although the importance of this effect
remains controversial \citep[][]{Mesinger2008}.  If present, this
radiative suppression of low mass galaxy-formation delays and extends
the reionization process, which though started by low mass galaxies,
must then be completed by relatively massive galaxies
\citep[e.g.][]{Iliev2007}. On the other hand, the semi-analytic
modelling of \citet[][]{Raicevic2011} indicates that photoionization
suppression of star formation in these low mass galaxies is unlikely
to significantly affect reionization, because of feedback effects on
the star formation efficiency implies that their contribution to
reionization is small.
The contribution of galaxies to the reionization of the IGM is
dependent on the star formation rate and initial stellar mass
function, but is also limited by the fraction of ionizing photons that
escape their host galaxies.  If the escape fraction is small, then
star formation had to be very efficient at high redshift in order to
reionize the Universe. The escape fraction is therefore a critical
parameter in studies of the connection between high redshift galaxy
formation and reionization.  Attempts to determine the escape fraction
have been dominated by direct observations of relatively low redshift
galaxies, and by numerical simulation.  Both observational
\citep[e.g.][]{Steidel2001, Fernandez2003, Shapley2006, Siana2007},
and theoretical \citep[e.g.][]{Razoumov2006, yajima2009, Gnedin2007a,
wise2009} estimates of the escape fraction at the Lyman-limit are
currently uncertain, with an expected range of $0.01\lsim f_{\rm
esc}\lsim1$.

In this paper we aim to utilise this new determination of the build up
of stellar mass at high redshift to constrain star formation scenarios
in high redshift galaxies, with particular attention to the possible
consequences of SNe feedback for the reionization of hydrogen. In
\S~\ref{model}, we describe an analytic model for the star formation
rate density function at high redshift based on the extended
Press-Schechter formalism \citep[][]{Lacey1993} with a simple physical
prescription for SNe feedback that suppresses star formation in low
mass galaxies.  We confront our model with the observed SFRD function
in \S~\ref{results} in order to constrain the starburst lifetime and
the stellar mass -- star formation rate relation, and show the
predictions of the model for the escape of ionizing photons from
star-forming galaxies in \S~\ref{Sfesc}. We investigate the
implications for the reionization history in
\S~\ref{reion}-\ref{contribution}, and conclude in
\S~\ref{conclusion}. In our numerical examples, we adopt the standard
set of cosmological parameters \citep{Komatsu2011}, with values of
$\Omega_{\rm b}=0.04$, $\Omega_{\rm m}=0.24$ and $\Omega_\Lambda=0.76$
for the density parameters of matter, baryon, and dark energy,
respectively, $h=0.73$, for the dimensionless Hubble constant, and
$\sigma_8=0.82$.

\section{Model}

\label{model}

The star formation rate in a galaxy halo of mass $M$ that turns a
fraction $f_\star$ of its disk mass $m_dM$ into stars over a time
$t_{\rm SF}$ is
\begin{equation}
\label{SFR}
SFR = 0.15 \mbox{M}_\odot \mbox{yr}^{-1}\left(\frac{m_{\rm
d}}{0.17}\right)\left(\frac{f_\star}{0.1}\right)\left(\frac{M}{10^8\mbox{M}_\odot}\right)\left(\frac{t_{\rm
SF}}{10^7\mbox{yr}}\right)^{-1} .
\end{equation}
We assume that major mergers trigger bursts of star formation, and
constrain the starburst lifetime required to reproduce the observed
star formation rate density function. The star formation rate density
function (i.e. galaxies per Mpc$^{-3}$ per unit of $SFR$) can be
estimated as
\begin{eqnarray}
\label{SFRD}
\nonumber
&&\hspace{-7mm}\Phi(SFR) = \\
&&\epsilon_{\rm duty}\left(\Delta M\left.t_{\rm H}\frac{dN^2_{\rm merge}}{dtd\Delta M}\right|_{M_1,\Delta M}\frac{dn}{dM}\right)\left(\frac{dSFR}{dM}\right)^{-1},
\end{eqnarray}
where $\epsilon_{\rm duty}$ is the fraction of the Hubble time
($t_{\rm H}$) over which each burst lasts, and $dn/dM$ is mass
function of dark matter halos \citep{Press1974,Sheth}. The rate of
major mergers ($dN_{\rm merge}/dt$) is calculated as the number of
halos per logarithm of mass $\Delta M$ per unit time that
merge\footnote{In addition to the excursion set approach of
\citep[][]{Lacey1993}, we have also computed merger rates using the
fitting formulae based on numerical simulations of
\citet[][]{Fakhouri2010}. We find consistent results using either
approach. } with a halo of mass $M_1$ to form a halo of mass $M$
\citep[][]{Lacey1993}.  We assign a 2:1 mass ratio to major mergers
(i.e. $M_1=\frac{2}{3}M$ and $\Delta M=M/3$).  The ionizing photon
rate from a starburst drops rapidly once the most massive stars fade
away on a timescale of $t_{\rm s}\sim3\times10^6$ years
\citep[][]{Barkana2001}. If the starburst lifetime $t_{\rm SF}$ is
much longer than $t_{\rm s}$, then the duty cycle associated with a
starburst is set by $t_{\rm SF}$ . However, if the starburst is shorter
than $t_{\rm s}$, the star formation remains visible for a minimum of
$t_{\rm s}$. Thus the duty-cycle is
\begin{equation} 
\epsilon_{\rm duty} = \frac{t_{\rm s} + t_{\rm SF}}{t_{\rm H}}.
\end{equation}
For comparison with observations we define 
\begin{equation}
\Psi(SFR) = \ln{10}\times SFR \times \Phi,
\end{equation}
which has units of Mpc$^{-3}$ per dex.

We expect that SNe feedback will alter the fraction of gas in a galaxy
that is turned into stars~\citep[e.g.][]{Dekel2003}. To determine the
mass and redshift dependence of $f_{\star}$ in the presence of SNe we
suppose that stars form with an efficiency $f_{\star}$ out of the gas
that collapses and cools within a dark matter halo and that a fraction
$F_{\rm SN}$ of each supernova energy output, $E_{\rm SN}$, heats the
galactic gas mechanically (allowing for some losses due to
cooling). The mechanical feedback will halt the star formation once
the cumulative energy returned to the gas by supernovae equals the
total thermal energy of gas at the virial velocity of the halo
\citep[e.g.][]{Wyithe2003a}. Hence, the limiting stellar mass is set
by the condition
\begin{equation}
\label{feedback}
\frac{M_{\star}}{w_{\rm SN}}E_{\rm SN}F_{\rm SN}f_{\rm t}f_{\rm d}=E_{\rm b}=\frac{1}{2}m_{\rm d}Mv_{\rm vir}^2.
\end{equation}
In this relation $E_{\rm b}$ is the binding energy in the halo,
$w_{\rm SN}$ is the mass in stars per supernova explosion, and the
total stellar mass is $M_\star=m_{\rm d}\,M\,f_{\rm \star,tot}$ where
$f_{\rm \star,tot}=N_{\rm merge}f_{\star}$ is the total fraction of
the gas that is converted to stars during major mergers, and $N_{\rm
merge}$ is the number of major mergers per Hubble time. The parameters
$f_{\rm t}$ and $f_{\rm d}$ denote the fraction of the SNe energy that
contributes because of the finite timescale of the SNe feedback 
or the disk scale height being smaller than the SNe bubble. These
terms are described in more detail below.

The ratio between the total mass in stars and dark matter is observed
to increase with halo mass as $(M_\star / M)\propto M^{0.5}$ for
$M_\star \la 3 \times 10^{10}$M$_\odot$, but is constant for larger
stellar masses \citep[][]{Kauffmann2003}. Thus, the star formation
efficiency within dwarf galaxies decreases towards low masses.  For
comparison with equation~(\ref{feedback}), a \citet[][]{Scalo1998}
mass function of stars has $w_{\rm SN} \sim 126$ M$_\odot$ per
supernova and $E_{\rm SN} =10^{51}$ ergs, and so we find that
$M_\star=3\times10^{10}$ M$_\odot$ and $v_{\rm c} \sim 175$ km/s
\citep[the typical value observed locally; see e.g.][]{Bell2001}
implies $f_{\rm \star,tot}\sim 0.1$ for a value of $F_{\rm SN} \sim
0.5$. Smaller galaxies have smaller values of
$f_{\star}$. Equation~(\ref{feedback}) indicates that
\begin{eqnarray}
\label{fstar}
\nonumber
&&\hspace{-7mm}f_{\star}=\\
&&\hspace{-7mm}\min\left[f_{\rm \star,max},\frac{0.008}{N_{\rm merge}}\left(\frac{M}{10^{10}\mbox{M}_\odot}\right)^{\frac{2}{3}}\left(\frac{1+z}{10}\right)\left(f_{\rm t}f_{\rm d}F_{\rm SN}\right)^{-1}\right].
\end{eqnarray}
We utilise equation~(\ref{fstar}) with equation~(\ref{SFRD}) as a
function of the parameters $t_{\rm SF}$ and $f_{\rm \star,max}$.

\subsection{Disk structure}

The effect of SNe feedback is dependent on the conditions of the interstellar medium (ISM)
gas.  We assume that the cold gas (out of which stars form) occupies a
self-gravitating exponential disk, with surface mass density
$\Sigma(r) = \Sigma_0 e^{-r/R_{\rm d}}$, where
\begin{equation}
\Sigma_0 = \frac{m_{\rm d} M}{2\pi R_{\rm d}^2}, 
\end{equation}
and $R_{\rm d}$ is the scale radius
\begin{equation}
R_{\rm d} = \frac{\lambda}{\sqrt{2}} R_{\rm vir},
\end{equation}
where $m_{\rm d}$ is the mass fraction of the disk relative to the halo
and $\lambda\sim 0.05$ is the spin parameter of the halo \citep[][]{Mo1998}.
The virial radius of a halo with mass $M_{\rm halo}$ is given by the
expression
\begin{equation}
\nonumber R_{\rm vir}= 0.784 h^{-1}\,\mbox{kpc} \left(\frac{M_{\rm halo}}{10^{8}M_{\odot}h}\right)^{\frac{1}{3}}
[\zeta(z)]^{-\frac{1}{3}}\left(\frac{1+z}{10}\right)^{-1},
\end{equation}
where $\zeta\equiv [(\Omega_{\rm m}/\Omega_{\rm
m}^z)(\Delta_c/18\pi^2)]$, $\Omega_{\rm m}^z \equiv
[1+(\Omega_\Lambda/\Omega_{\rm m})(1+z)^{-3}]^{-1}$,
$\Delta_c=18\pi^2+82d-39d^2$, and $d=\Omega_{\rm m}^z-1$ \citep[see
equations~22--25 in][for more details]{Barkana2001}. The scale height
of the disk at radius $r$ is
\begin{equation}
H=\frac{c_{\rm s}^2}{\pi G \Sigma(r)}.
\end{equation}
We adopt the density in the mid plane at the scale radius, within
which half the gas is contained, as representative of the density of
the ISM. At $r=R_{\rm d}$ the scale height is
\begin{eqnarray}
\nonumber H&=&\frac{2 c_{\rm s}^2 R_{\rm d}^2 \times 2.71}{G m_{\rm d}
M}\\ \nonumber &=& 0.034\,\mbox{kpc}
\left(\frac{\lambda}{0.05}\right)^{2} \left(\frac{m_{\rm
d}}{0.17}\right)^{-1}
\left(\frac{M}{10^8\mbox{M}_\odot}\right)^{-1/3}\\ &&\hspace{15mm}
\times\left(\frac{1+z}{10}\right)^{-2}\left(\frac{c_{\rm
s}}{10\mbox{km/s}}\right)^2.
\end{eqnarray}
The number density of particles in the mid plane at $r=R_d$ is
\begin{equation}
n_{\rm p} = \frac{G(m_{\rm d}M)^2}{8\pi m_{\rm p} c_{\rm s}^2 R_{\rm
d}^4 \times 2.71^2},
\end{equation}
where $m_{\rm p}$ is the mass density of
particles. 

We note that if the gas disk becomes stable to fragmentation
at a radius beyond which there is a significant fraction of gas by
mass, then the half mass radius of the stellar disk may not equal the
scale radius of the gas-disk. However, \citet[][]{WyitheL2011} find
that the disk becomes stable (based on Toomre--$Q$ criterion) only at
3--4 scale radii.

\subsection{Supernova evacuation of the ISM}

\citet[][]{Clarke2002} presented a simple analytic model for the
effect of supernovae on the interstellar medium which we apply to high
redshift galaxies. In this model, clusters of $N_{\rm e}$ SNe produce
super-bubbles in the ISM with a radius $R_{\rm e}$ at which the
super-bubble comes into pressure balance with the ISM. This radius can
be found by approximating $R_{\rm e}$ as the radius within which the
thermal energy of the ISM equals the mechanical energy of the SNe
cluster. The timescale associated with the evacuation of a super bubble in the ISM
 by a SNe cluster is $t_{\rm e}=4\times10^7$
years, corresponding to the lifetime of the lowest mass SNe progenitor.
 The evacuation radius for a cluster of $N_{\rm e}$ SNe, each
with energy output $E_{\rm SN}$ within an ISM of sound speed $c_{\rm
s}$ is therefore
\begin{equation}
R_{\rm e}=\left(\frac{2N_{\rm e} E_{\rm SN}}{2\pi m_{\rm p} n_{\rm p} c_{\rm s}^2}\right)^{\frac{1}{3}},
\end{equation} 
yielding
\begin{eqnarray}
\label{re}
\nonumber
R_{e} &=&0.08 \,\mbox{kpc} \left(\frac{N_{\rm e}}{10}\right)^{\frac{1}{3}} \left(\frac{E_{\rm SN}}{10^{51}\mbox{erg}}\right)^{\frac{1}{3}}  \left(\frac{\lambda}{0.05}\right)^{\frac{4}{3}} \left(\frac{m_{\rm d}}{0.17}\right)^{-\frac{2}{3}} \\
&&\hspace{5mm}\times\left(\frac{M}{10^8\mbox{M}_\odot}\right)^{-\frac{2}{9}} \left(\frac{1+z}{10}\right)^{-\frac{4}{3}}.
\end{eqnarray}

The derived value of $R_{\rm e}$ ignores radiative losses of the
super-bubble before it comes into pressure equilibrium with the
ISM. \citet[][]{Clarke2002} evaluated the validity of this assumption
by noting that in galaxies like the Milky-Way, the radius at which
cooling becomes important is larger than the radius at which the
super-bubble comes into pressure equilibrium. Of importance here is
the clustering of SNe, which concentrates the mechanical output into
small regions of the ISM. At high redshift, cooling is expected to be
more efficient in the much denser ISM, although this will be offset by
the lower gas metallicity. \citet[][]{MacLow1988} derived the cooling
radius to be
\begin{equation}
R_{\rm c}=350\mbox{pc}\left(\frac{L}{10^{38}\mbox{erg/s}}\right)^{4/11}n_{\rm p}^{-7/11}\zeta^{-27/22},
\end{equation}
where $L$ is the mechanical luminosity of the SNe cluster and
$\zeta=Z/Z_\odot$ is the ISM metallicity in units of the solar value.
Taking $L=N_{\rm e}E_{\rm SN}/t_{\rm e}$,
\begin{eqnarray}
\label{rc}
\nonumber
R_{\rm c} &=&0.12 \,\mbox{kpc} \left(\frac{N_{\rm e}}{10}\right)^{\frac{4}{11}} \left(\frac{E_{\rm SN}}{10^{51}\mbox{erg}}\right)^{\frac{4}{11}}  \left(\frac{t_{e}}{4\times10^7\mbox{yr}}\right)^{-\frac{4}{11}} \\
\nonumber
&\times&\left(\frac{\lambda}{0.05}\right)^{\frac{28}{11}} \left(\frac{m_{\rm d}}{0.17}\right)^{-\frac{14}{11}}\left(\frac{c_{\rm s}}{10\mbox{km/s}}\right)^{-\frac{14}{11}}  \\
&\times&\left(\frac{M}{10^{8}\mbox{M}_\odot}\right)^{-\frac{14}{33}} \left(\frac{1+z}{10}\right)^{-\frac{28}{11}}\left(\frac{\zeta}{0.05}\right)^{-\frac{27}{22}}.
\end{eqnarray}
Comparing equation~(\ref{rc}) with equation~(\ref{re}) for $R_{\rm e}$
we find that the assumption of adiabatic expansion is valid in the low
mass galaxies thought to drive reionization. We therefore adopt
equation~(\ref{re}) for $R_{\rm e}$ in the remainder of this paper.

In the limit where SNe evacuated regions are smaller than the scale
height of the disk, and the starburst lifetime $t_{\rm SF}$ is much
larger than the gas evacuation timescale $t_{\rm e}$, the fraction
$F_{\rm SN}$ of the SNe energy may be used in feedback suppressing
subsequent star formation. However, if the SNe evacuated regions break
out of the disk, or $t_{\rm SF}<t_{\rm e}$, not all of the energy will
be available for feedback. Based on the ISM porosity model of
\citet[][]{Clarke2002}, a fraction $f_{\rm d}=2H/R_{\rm e}$ of the SNe
energy goes to increasing the ISM porosity for disks where $R_{\rm
e}>H$. In this case we find
\begin{eqnarray}
\nonumber
f_{\rm d}&=&0.85 \left(\frac{N_{\rm e}}{10}\right)^{-\frac{1}{3}} \left(\frac{E_{\rm SN}}{10^{51}\mbox{erg}}\right)^{-\frac{1}{3}}  \left(\frac{\lambda}{0.05}\right)^{\frac{2}{3}} \left(\frac{m_{\rm d}}{0.17}\right)^{-\frac{1}{3}} \\
&\times&\hspace{5mm}\left(\frac{M}{10^8\mbox{M}_\odot}\right)^{-\frac{1}{9}} \left(\frac{1+z}{10}\right)^{-\frac{2}{3}}\left(\frac{c_{\rm s}}{10\mbox{km/s}}\right)^2,
\end{eqnarray}
as long as $f_{\rm d}<1$ and $f_{\rm d}=1$ otherwise. Similarly, in
cases where $t_{\rm SF}<t_{\rm e}\sim 4\times10^7$ yrs, only
\begin{equation}
f_{\rm t}\equiv(t_{\rm SF}/t_{\rm e})^2
\end{equation}
of the overall SNe energy output is generated by the time the
starburst concludes. The quadratic dependence on time arises because
the number of bubbles produced grows in proportion to time, while the
maximum size of a bubble at time $t<t_{\rm e}$ is also proportional to
time \citep[][]{Oey1997}. In cases where $t_{\rm SF}>t_{\rm e}$ we
have $f_{\rm t}=1$.

In the next section we fit our model to the recent data of
\citet[][]{Smit2012} in order to constrain $t_{\rm SF}$ and
$f_{\star}$.

\section{Results}
\label{results}

\begin{figure*}
\begin{center}
\vspace{3mm}
\includegraphics[width=17.5cm]{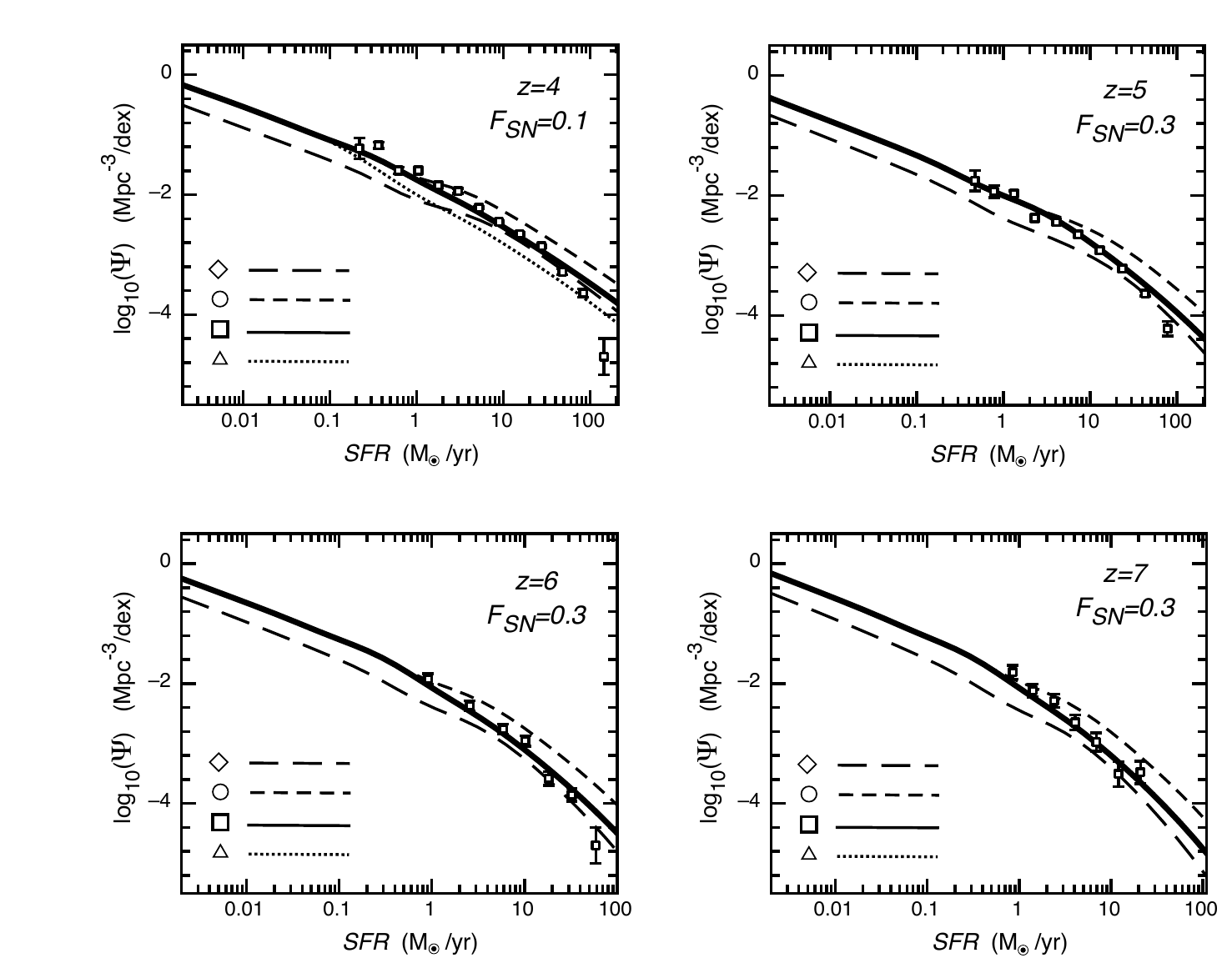}
\caption{\label{fig1}Comparison between the observed and modelled SFRD
function (plotted as $\Psi = \ln{10}\times SFR\times\Phi$). The four panels
show results for different redshift values.  In each panel, the four
curves correspond to a different choice of model parameters $t_{\rm
SF}$ and $f_{\rm \star,max}$, labeled by the symbols in
Figure~\ref{fig2}. The thick solid lines represent values close to the
best fit. }
\end{center}
\end{figure*}

\begin{figure*}
\begin{center}
\vspace{3mm}
\includegraphics[width=15cm]{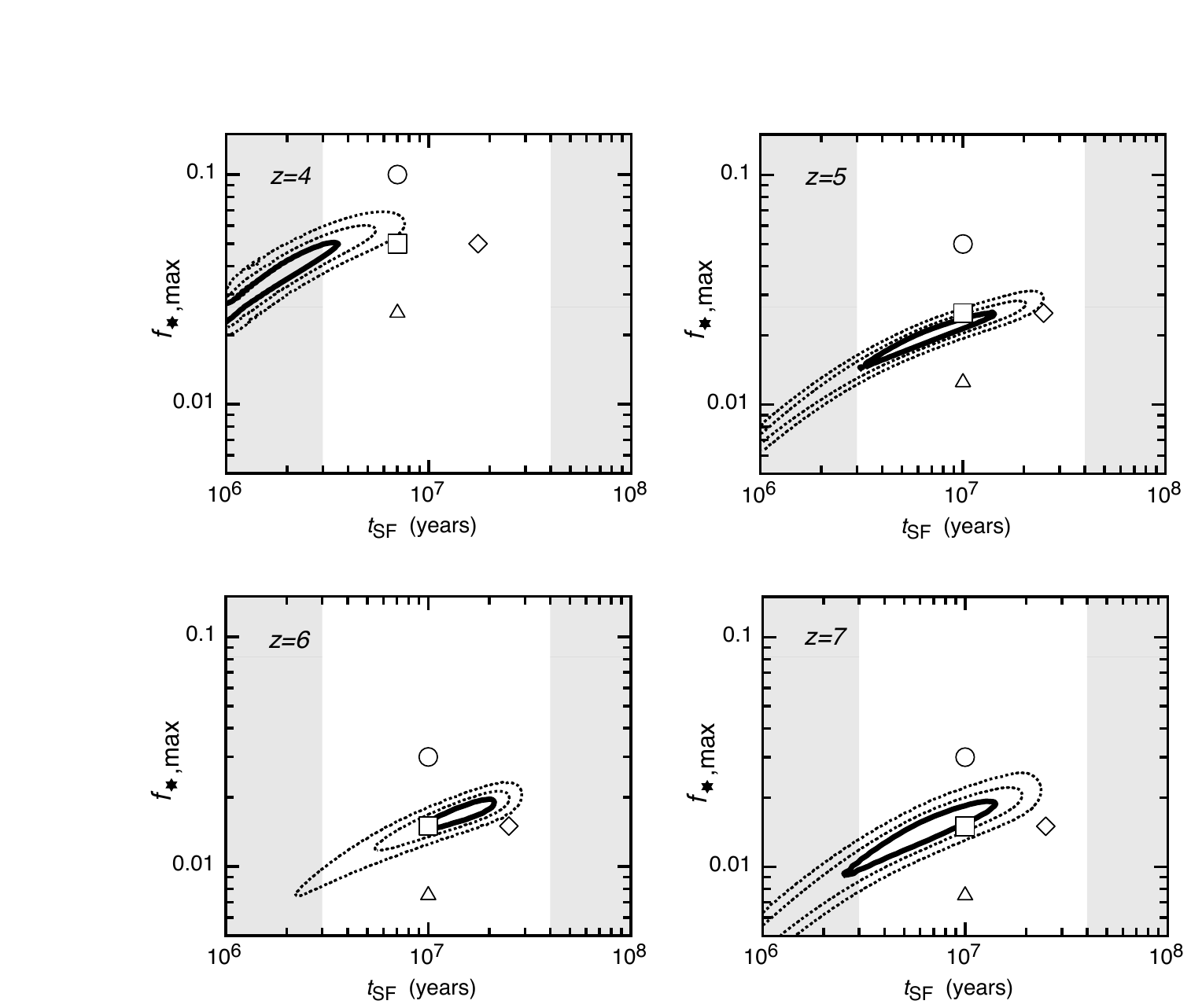}
\caption{\label{fig2}Constraints on the model parameters $f_{\rm
\star,max}$ and $t_{\rm SF}$ at four different redshifts (constraints
are independent at each redshift).  In each case, three contours are
shown corresponding to differences in $\chi^2$ relative to the
best-fitting model of $\Delta\chi^2 = \chi^2 - \chi^2_{\rm min} = 1$,
2.71 and 6.63. Projections of these contours on to the axes provide
the 68.3, 90 and 99 per cent confidence intervals on individual
parameter values. The vertical grey regions represent time-scales
longer/shorter than the lifetime of the highest/lowest mass SNe
progenitor ($3\times10^6$yr$/4\times10^7$yr). }
\end{center}
\end{figure*}

\begin{figure*}
\begin{center}
\vspace{3mm}
\includegraphics[width=15cm]{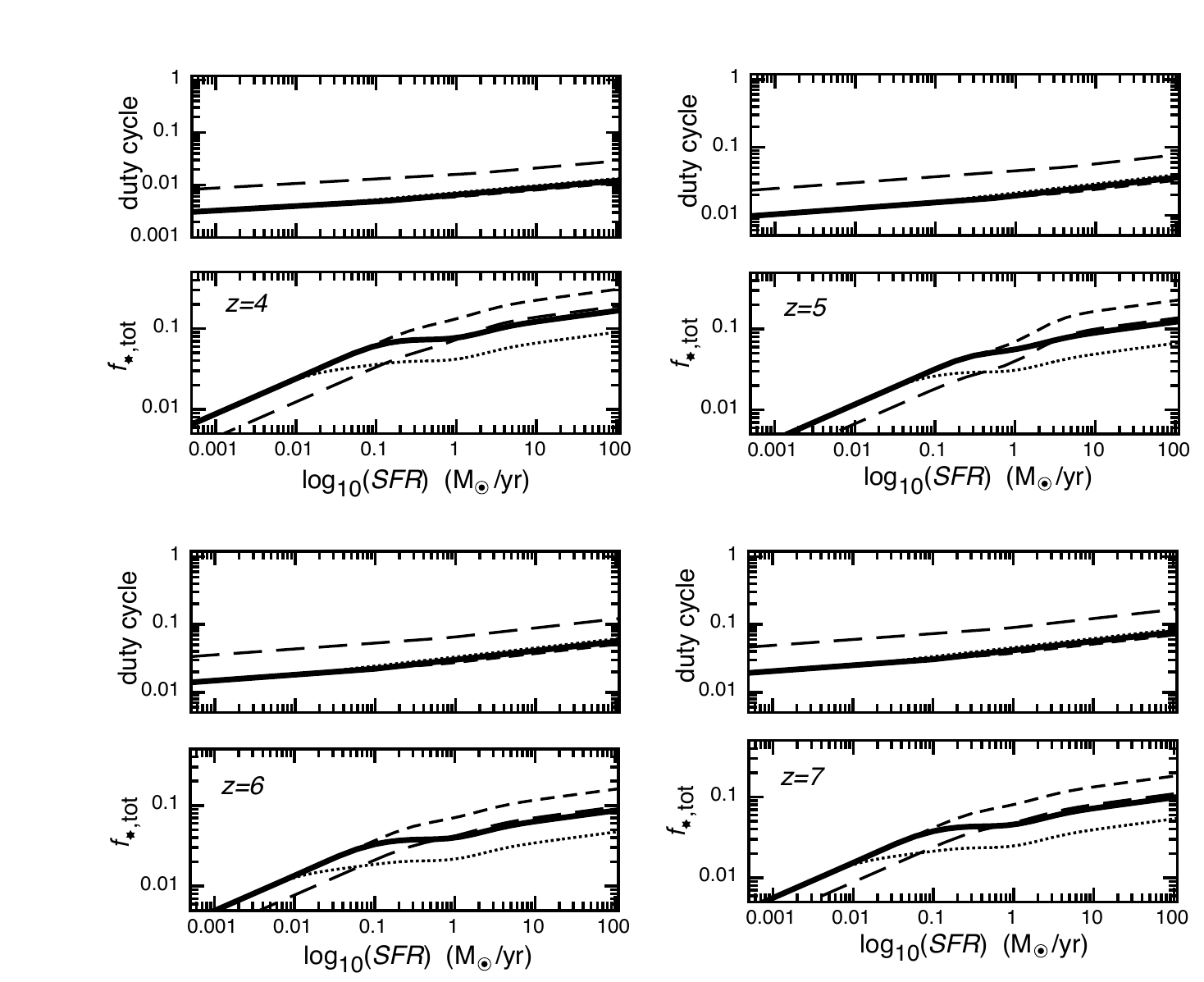}
\caption{\label{fig3}The values of total star formation efficiency
$f_{\rm \star,tot}$ (i.e. the sum of $f_{\rm \star}$ over all
mergers), and the overall duty-cycle (i.e. the fraction of a Hubble
time during which a galaxy is starbursting) as a function of $SFR$.
The four curves shown correspond to the SFRD functions shown in
Figure~\ref{fig1}, with model parameters $t_{\rm SF}$ and $f_{\rm
\star,max}$ designated by the symbols in Figure~\ref{fig2}. }
\end{center}
\end{figure*}

Observations of the star formation rate density function (SFRD) are
plotted in Figure~\ref{fig1} for four different redshifts $z\sim4$,
5, 6 and 7. From these observations we constrain the two free
parameters of our star formation model, $f_{\rm \star,max}$ and
$t_{\rm SF}$. We use the model to calculate SFRD functions for
combinations of these parameters and calculate the $\chi^2$ of the
model as
\begin{eqnarray} 
\nonumber
\chi^2(f_{\rm \star,max},t_{\rm SF})&=&\\
&&\hspace{-30mm}\sum_{i=0}^{N_{\rm obs}}\left(\frac{\log{\Psi(SFR_i,f_{\rm \star,max},t_{\rm SF},z)}-\log{\Psi_{\rm obs}(SFR_i,z)}}{\sigma_{SFR}(SFR_i,z)}\right)^2.
\end{eqnarray}
Here $\Psi_{\rm obs}(SFR_i,f_{\rm \star,max},t_{\rm SF},z)$ is the
observed star formation rate density measured at redshift $z$, with
uncertainty in dex of $\sigma_{SFR}(SFR_i)$. In calculating
likelihoods at $z\sim4$ and $z\sim5$ we increased the quoted error
bars by factors of 3 and 2 respectively in order to obtain a reduced
$\chi^2$ of order unity. The SFRD function is sensitive to the value
of $F_{\rm SN}$, 
and we therefore integrate the likelihood over a range of values
uniformly distributed between $-1<\log_{10}{F_{\rm SN}}<0$
\begin{equation}
\mathcal{L}(f_{\rm \star,max},t_{\rm SF})\propto\int_{-1}^0 d(\log_{\rm 10}{F_{\rm SN}}) e^{-\chi^2/2}.
\end{equation}

We note that the relation between $SFR$ and $M$ in
equation~(\ref{SFR}) is not perfect. As part of our comparison with
observations, and to account for scatter in this relationship, we
convolve the predicted SFRD function equation~(\ref{SFRD}) with a
Gaussian of width $0.5$ dex in $SFR$.  An intrinsic scatter of 0.5 dex
is motivated by the scatter in stellar mass at constant $SFR$ found by
\citet[][]{Gonzalez2011}. However, in addition we find that the value
of 0.5 dex provides the best statistical fit to the observations. Our
qualitative results are not sensitive to the choice of this scatter,
however.  

\subsection{Parameter constraints}

Constraints on $f_{\rm \star,max}$ and $t_{\rm SF}$ for this model are
shown in Figure~\ref{fig2}. We note that these constraints are formal
values assuming the particular model (for example, specifying the
scale of a major merger). They are therefore indicative of the
parameters describing the star formation model, whereas the overall
uncertainty on starburst lifetime and efficiency may be larger than
the ranges shown.  We find that the shape of the SFRD function
requires starburst durations of a few tens of Myr
at $z\sim5$, $z\sim6$ and $z\sim7$, with a few percent of the gas
turned into stars per burst. For comparison the left and right hand
vertical grey regions represent times smaller than the lifetime of the
most massive stars ($t_{\rm s}\sim3\times10^6$ years), and times in
excess of the lifetime of the least massive stars that produce SNe
respectively. Our results therefore indicate that star formation in
high redshift galaxies is terminated on the same timescale as feedback
from SNe can be produced.  The constraints can be understood
qualitatively. First, larger values of $f_{\rm \star,max}$ at fixed
$t_{\rm SF}$ lead to smaller values of halo mass at fixed $SFR$, and
hence a larger SFRD function. Similarly, a longer starburst lifetime
and hence duty cycle, implies that smaller halo masses are needed for
a given value of $SFR$.  Figure~\ref{fig1} shows the comparison
between observed and modelled SFRD functions. The four curves shown
correspond to model parameters $t_{\rm SF}$ and $f_{\rm \star,max}$
labeled by the symbols in Figure~\ref{fig2}. We show the case of
$F_{\rm SN}=0.3$, except at $z\sim4$ where the lower value of $F_{\rm
SN}=0.1$ produces a better fit.  The thick solid lines provide the
best fit to the observational data. The agreement with the data at
multiple redshifts ($z=5,6$ and 7) is impressive since the values of
$f_{\rm \star,max}$ and $t_{\rm SF}$ were kept fixed.  We find that
the SFRD function should continue to increase to much fainter levels
than currently observed.

The parameters $f_{\rm \star,max}$ and $t_{\rm SF}$ refer to single
bursts, whereas our model includes multiple bursts at the rate of
major mergers. We therefore calculate the total star formation
efficiency $f_{\rm \star,tot}=N_{\rm merge}f_\star$ (i.e. the sum of
$f_{\rm \star}$ over all mergers), as well as the overall duty-cycle
$N_{\rm merge}t_{\rm SF}/t_{\rm H}$.  These quantities are plotted in
Figure~\ref{fig3} based on our model with parameter choices
corresponding to the examples in Figure~\ref{fig1}. We find
duty-cycles of a few percent, with higher duty-cycles at higher
redshift reflecting the increased ratio between the lifetime of
massive stars and the age of the Universe. We find that $\sim5-10\%$ of
the gas forms stars in bright galaxies of $SFR\sim1-100$M$_\odot$ per year,
with lower fractions down to a percent in fainter galaxies.

\subsection{The halo masses of star-forming galaxies}

\begin{figure*}
\begin{center}
\vspace{3mm} \includegraphics[width=15cm]{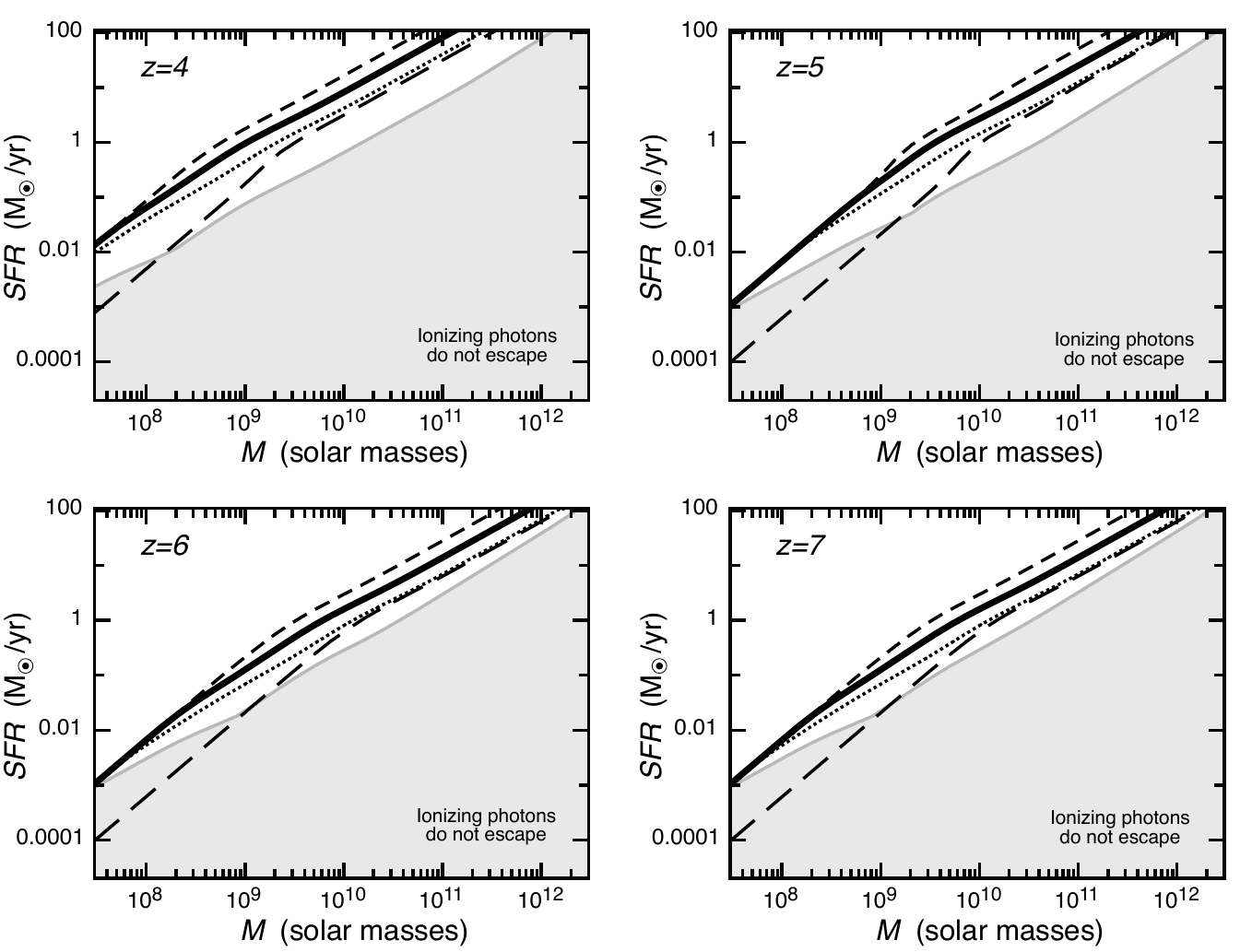}
\caption{\label{fig4} Relation between $SFR$ and halo mass $M$ at four
different redshifts based on our model with parameter choices
corresponding to the examples in Figure~\ref{fig1}. Also shown is the
critical star formation rate required to achieve porosity equal to unity
($SFR_{\rm crit}$) as a function of halo mass. Regions of the $M-SFR$
plane where the porosity $<1$, and hence ionizing photons do not
escape, are shaded grey.}
\end{center}
\end{figure*}

In Figure~\ref{fig4} we show the relation between SFR and halo mass
$M$ for galaxies based on our model with parameter choices
corresponding to the examples in Figure~\ref{fig1}. Galaxies observed
to have $SFR\sim1$M$_\odot/$yr reside in halos of
$M\sim10^{10}$M$_\odot$, a result consistent with previous work
\citep[e.g.][]{Trenti2010,Munoz2011}. Halos thought to host the
smaller galaxies at the hydrogen cooling limit ($M\sim10^{7.5}$M$_\odot$)
are predicted to have low star formation rates of
$SFR\sim10^{-3}$M$_\odot/$yr.

\subsection{The stellar masses of star-forming galaxies}

\begin{figure*}
\begin{center}
\vspace{3mm}
\includegraphics[width=15cm]{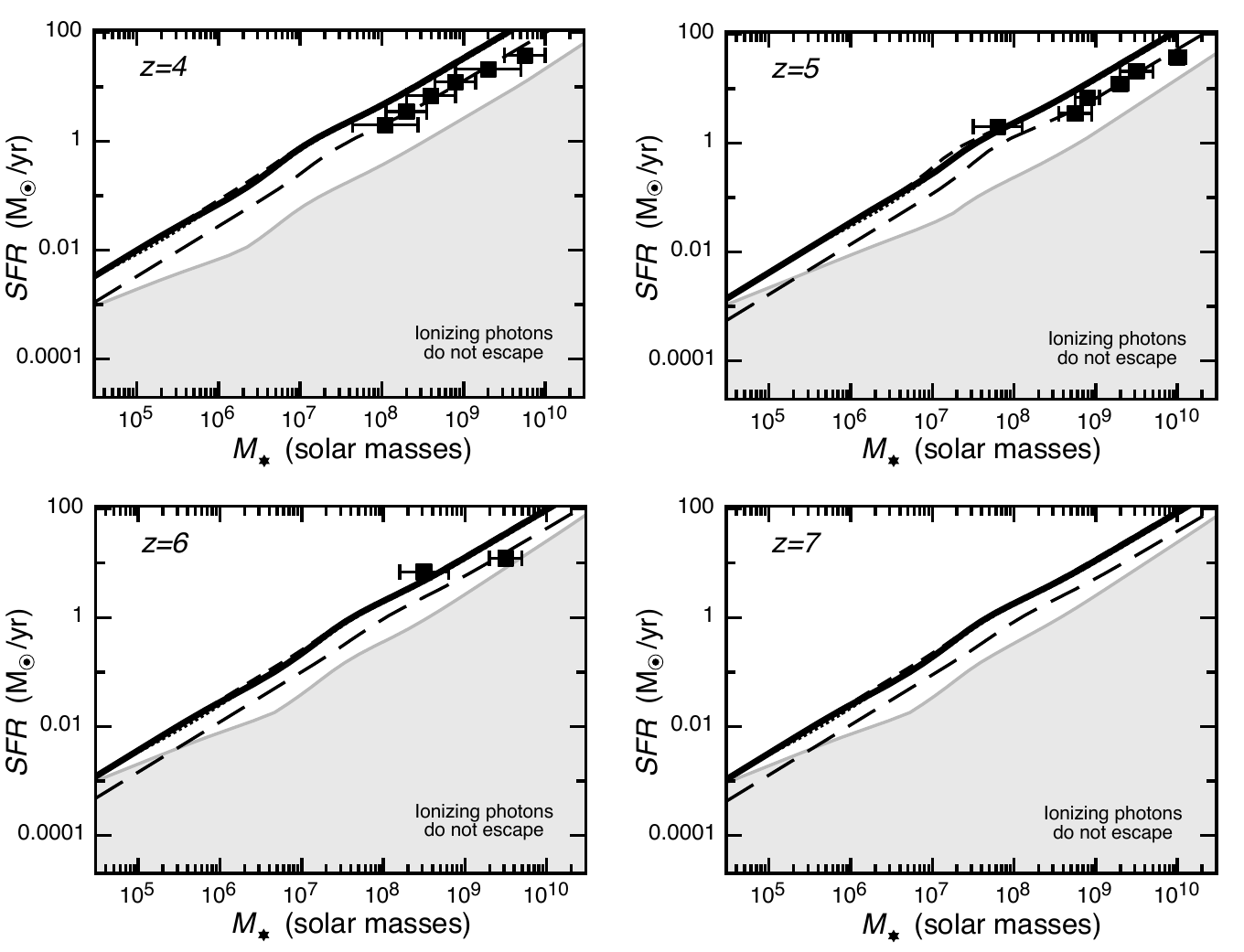}
\caption{\label{fig5} Relation between $SFR$ and stellar mass
$M_\star$ at four different redshifts based on our model with
parameter choices corresponding to the examples in
Figure~\ref{fig1}. Also shown is the critical star formation rate
required to achieve porosity of unity ($SFR_{\rm crit}$) as a function
of halo mass. Regions of the $M_\star-SFR$ plane where the porosity
$<1$, and hence ionizing photons do not escape are shaded grey. The
points show the observed relations between stellar masses and star
formation rates. We have corrected the star formation rates in
\citet[][]{Gonzalez2011} for extinction using the methods outlined in
\citet[][]{Smit2012}, based on relations in \citet[][]{Meurer1999} and
\citet[][]{Bouwens2012}.}
\end{center}
\end{figure*}

In Figure~\ref{fig5} we show the relation between SFR and stellar
mass $M_\star=m_{\rm d}f_{\rm \star,tot}M$ for galaxies based on our
model with parameter choices corresponding to the examples in
Figure~\ref{fig1}. $SFR\sim1$M$_\odot/$yr characterise galaxies with
$M_\star\sim10^{8}$M$_\odot$. Our model yields a close to linear
relation between stellar mass and star formation rate, in good
agreement with simulations \citep[e.g.][]{Jaacks2012,Salvaterra2011}.

In Figure~\ref{fig5} we also show data points representing the mean
relation between observed star formation rates and stellar masses. The
relation between stellar masses and extinction-uncorrected star
formation rates at $z\sim4$, 5 and 6, was presented in
\citet[][]{Gonzalez2011}. We have corrected these star formation rates
for extinction using the methods outlined in \citet[][]{Smit2012},
based on relations in \citet[][]{Meurer1999} and
\citet[][]{Bouwens2012}. We find that our predicted $SFR-M_\star$
relation is in agreement with the observations. In
particular, we note that while the relation is fairly insensitive to
the value of $f_{\rm \star,max}$, larger values of $t_{\rm SF}$ than
suggested by our modelling of the SFRD function would imply a star
formation rate at fixed stellar mass that is lower than required by
the observations.

\begin{figure*}
\begin{center}
\vspace{3mm}
\includegraphics[width=15cm]{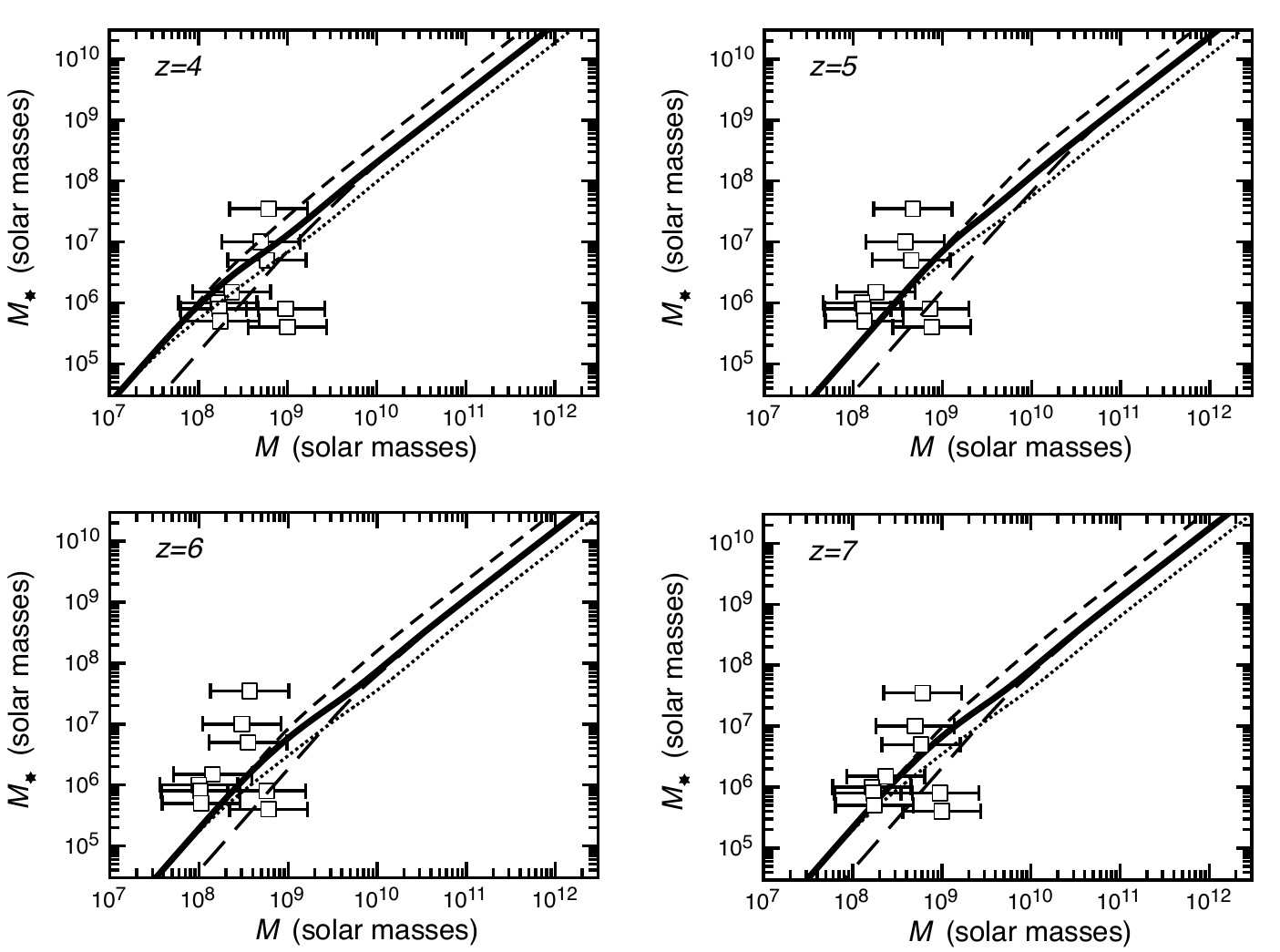}
\caption{\label{fig6} Relation between halo mass $M$ and stellar mass
$M_\star$ at four different redshifts based on our model with
parameter choices corresponding to the examples in Figure~\ref{fig1}.
The data points are based on the relation between stellar mass and
maximum circular velocity for Milky-Way dwarf spheroidals
\citep[][]{Boylan2012}. To convert to
halo mass we adopt the values of virial velocity at time of collapse
from \citet[][]{Boylan2012}, and calculate the corresponding halo mass
at redshift $z$.}
\end{center}
\end{figure*}

In Figure~\ref{fig6} we show the relations between the halo mass and
stellar mass. The stellar to halo mass ratio is constant at high
masses ($M\gtrsim10^{9}$M$_\odot$), but steeper towards low masses
owing to the SNe feedback lowering the star formation efficiency in
the model. Recently, \citet[][]{Rocha2012} suggested that the
Milky-Way dwarf spheroidals appear to have had their star formation
quenched at the time of infall into the Milky-Way, and that this time
of infall was typically $\sim7-10$ Gyr ago. If true this implies that
the Milky-Way dwarf spheroidals represent fossil records of the
star-forming galaxies during reionization, and should have stellar
masses described by our model. To make this comparison we therefore
over-plot the observed relation for Milky-Way dwarf spheroidals
\citep[][]{Boylan2012}. To convert to
halo mass we adopt the values of virial velocity at time of collapse
from \citet[][]{Boylan2012}, and calculate the corresponding halo mass
at each redshift $z$. We find good agreement between the observed
relation (although the scatter is large), indicating that if the dwarf
spheroidals are old galaxies formed at around the end of reionization,
then they would have the stellar to halo mass ratio predicted by our
SNe feedback limited model. Combined with the correct prediction of
the stellar mass for star-forming galaxies during reionization, this
implies that our model correctly describes the stellar mass to halo
mass ratio in the range $10^5$M$_\odot\lsim
M_\star\lsim10^{10}$M$_\odot$.

\section{The Escape fraction of ionizing photons}
\label{Sfesc}

The contribution of galaxies to reionization is dependent on the star
formation rate and initial stellar mass function, but is also limited
by the fraction of ionizing photons that escape their host galaxies.
If the escape fraction is small, then star formation had to be very
efficient at high redshift in order to reionize the Universe. The
escape fraction is therefore a critical parameter in studies linking
high redshift galaxy formation to reionization. In this section we
discuss the implications of our model for the ionizing photon escape
fraction.

Attempts to determine the ionizing photon escape fraction have been
dominated by direct observations of relatively low redshift galaxies,
and by numerical simulation.  Observational estimates of the escape
fraction at the Lyman-limit are currently uncertain, with no confident
detections at $0<z<1$, and only some detections at $z>3$.  At
redshifts $z\sim1-3$, observations have suggested a broad range of
values for $f_{\rm esc}$, from a few percent to $\gtrsim 20\%$
\citep[e.g.]{Steidel2001, Fernandez2003, Shapley2006,
Siana2007}. \citet{Inoue2006} have examined the evolution of the
escape fraction in the redshift range $z=0-6$ using both direct
observations of the escape fraction and values that they derive from
measurements of the ionizing background.  They find that the escape
fraction evolves from $f_{\rm esc} \sim 1-10\%$, increasing towards
high redshift.

Theoretically, \citet{Razoumov2006} used galaxy formation simulations
incorporating high-resolution 3-D radiative transfer to show that the
escape fraction evolves from $f_{\rm esc}\sim1$--$2\%$ at $z=2.39$ to
$f_{\rm esc}\sim6$--$10\%$ at $z=3.6$.  In agreement with
\citet{Fujita2003}, \citet{Razoumov2006} \citep[see
also][]{yajima2009} find that increased supernova feedback at higher
redshift expels gas from the vicinity of starbursting regions,
creating tunnels in the galaxy through which ionizing photons can
escape into the IGM.  Numerical simulations \citep{Gnedin2007a} have
predicted a value for $f_{\rm esc}$ between $1$ and $3\%$, for halos of
mass $M\gtrsim5\times10^{10}M_{\odot}$, over the redshift range
$3<z<9$.  This very low efficiency of reionization would have profound
implications for the reionization history.  In addition to a small
escape fraction in massive galaxies, \citet{Gnedin2007a} further
predict that halos with $M\lsim 5\times10^{10}M_{\odot}$ have an
escape fraction that is negligibly small. However, more recently
\citet{wise2009} have used a large suite of simulations to show that
the time averaged escape fraction for dwarf galaxies is expected to be
large ($>25\%$). Since dwarf galaxies are thought to dominate the
ionizing flux, resolution of this issue is of primary importance for
studies of reionization. Overall, the escape fraction is predicted to
span a very broad range, $0.01\lsim f_{\rm esc} \lsim 1$. This broad
range may be explained by inhomogeneities in the hydrogen distribution
within galaxies \citep[][]{Dove2000,Fernandez2011}, or by variations
in viewing angle \citep[][]{Wood2000}.

Within our formalism we follow \citet[][]{Clarke2002} who proposed
that galaxies with a sufficient star formation rate to generate a
porosity greater than unity had an escape fraction for ionizing
photons that is of order unity, whereas galaxies with insufficient
start formation have a negligible escape fraction. The critical star
formation rate required to achieve porosity of unity is
\citep[][]{Clarke2002}
\begin{equation}
\label{sfrcrit}
SFR_{\rm crit}=0.15 \mbox{M}_\odot \mbox{yr}^{-1}\left(f_{\rm t}f_{\rm
d}\right)^{-1}\left(\frac{m_{\rm d}M}{10^{10}\mbox{M}_\odot}\right)
\left(\frac{c_{\rm s}}{10\mbox{km/s}}\right)^2.
\end{equation}

In Figures~\ref{fig4} and \ref{fig5} we plot the curves corresponding
to $SFR_{\rm crit}$ as a function of halo mass and stellar mass,
respectively. We shade the areas below the line for clarity to
indicate the regions in the $M-SFR$ and $M_\star-SFR$ planes where the
porosity $<1$ and hence ionizing photons do not escape. The model
predicts that galaxies with halo masses in excess of
$M\sim10^8$M$_\odot$ (corresponding to $M_\star>10^{5}$M$_\odot$)
have a sufficiently large $SFR$ to potentially contribute to reionization. 

\begin{figure*}
\begin{center}
\vspace{3mm}
\includegraphics[width=15cm]{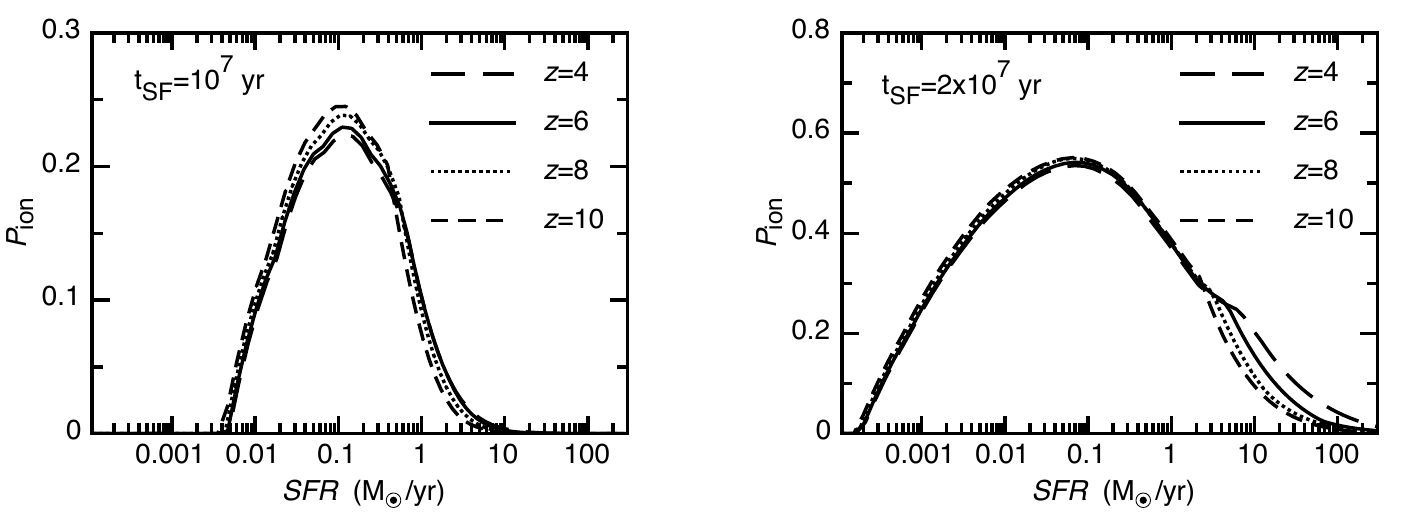}
\caption{\label{fig7} Curves of $P_{\rm ion}$ as a function of
$SFR$. The left hand panel shows the of $t_{\rm SF}=10^7$ yr
corresponding to the best fit example in Figure~\ref{fig1}. The right
hand panel shows the larger value of $t_{\rm SF}=2\times10^7$ yr. In
each case curves are shown for $z\sim4$, 6, 8 and 10. }
\end{center}
\end{figure*}

In the model of \citet[][]{Clarke2002} galaxies
with $SFR>SFR_{\rm crit}$ attain porosity of unity after a time
$t_{Q}=t_{\rm e}(SFR/SFR_{\rm crit})^{-1/2}$. The ionizing photon rate
from a starburst drops rapidly once the most massive stars fade away
on a timescale of $t_{\rm s}\sim3\times10^6$ years
\citep[][]{Barkana2001}.  Since ionizing photons are produced for a
time $t_{\rm SF}+t_{\rm s}$ but can only escape the galaxy after a
time $t_{Q}$, the time during which the galaxy with $SFR>SFR_{\rm
crit}$ emits photons into the IGM is
\begin{eqnarray}
\nonumber 
t_{\rm ion}&=&t_{\rm SF}+t_{\rm s}-t_Q\\
&=&t_{\rm SF}+t_{\rm s}\left(1-\left(\frac{t_{\rm e}}{t_{\rm s}}\right)\left(\frac{SFR}{SFR_{\rm crit}}\right)^{-1/2}\right).
\end{eqnarray}
For small values of $t_{\rm SB}<t_{e}$, $Q$ may never reach unity for
some values of $SFR$, even if $SFR>SFR_{\rm crit}$. To estimate
$t_{\rm ion}$ we must therefore integrate over the distribution for
$SFR$ at fixed halo mass, which was assumed to have a scatter of
0.5dex. Thus we calculate
\begin{equation}
\label{tion}
\langle t_{\rm ion}\rangle = \int d\Delta_{SFR}\, t_{\rm ion}(SFR\times10^{\Delta_{SFR}}) e^{-\frac{(\Delta_{SFR})^2}{2\times0.5^2}}.
\end{equation}
We note that since $\langle t_{\rm ion}\rangle<t_{\rm SF}+t_{\rm s}$,
we expect a wide range of observed escape fractions for star forming
galaxies depending on whether or not the starburst is being observed
before or after $t_Q$. The galactic porosity model of
\citep[][]{Clarke2002} postulated that the escape fraction is
negligible at $t<t_Q$ and of order unity at $t>t_Q$. Our model
provides the probability $P_{\rm ion}$ that a galaxy with a particular
$SFR$ should be observed with a non-zero escape fraction for ionizing
radiation, calculated as the fraction of time that the galaxy is
observed to be star-forming and for which $t>t_{Q}$,
\begin{equation}
P_{\rm ion} = \frac{\langle t_{\rm ion}\rangle}{t_{SF}+t_{\rm s}}.
\end{equation}
Curves of $P_{\rm ion}$ are plotted as a function of $SFR$ in
Figure~\ref{fig7}. The left hand panel shows the best fit case of
$t_{\rm SF}=10^7$ yr, and the right hand panel $t_{\rm
SF}=2\times10^7$ yr. Four redshifts are shown. Our model predicts that
large escape fractions should be rare for both low SFR and high SFR
galaxies. However, we expect approximately a half of star-forming
galaxies with $SFR\sim0.1-1$M$_\odot$/yr to have a significant escape
fraction. Thus, our model provides a natural explanation for the wide
range of conclusions regarding observations of the escape fraction of
ionizing photons from star-forming galaxies.
 
Next we investigate the implications of this model for the
reionization history.

\section{Implications for the Reionization History}
\label{reion}

We continue with a semi-analytic calculation of the reionization
history of the IGM based on the star formation model presented in
\S~\ref{model}. The basis for our model of reionization is the excess
ionization rate over the recombination rate for hydrogen in an
inhomogeneous IGM.  \citet{Miralda2000} presented a model that allows
the calculation of an effective recombination rate in an inhomogeneous
universe by assuming a maximum overdensity ($\Delta_{\rm c}$)
penetrated by ionizing photons within HII regions. The model assumes
that reionization progresses rapidly through islands of lower density
prior to the overlap of individual cosmological ionized
regions. Following the overlap epoch, the remaining regions of high
density are gradually ionized. It is therefore hypothesized that at
any time, regions with gas below some critical over density
$\Delta_{\rm i}\equiv {\rho_{i}}/{\langle\rho\rangle}$ are highly
ionized while regions of higher density are not.  The fraction of mass
in regions with overdensity below $\Delta_{\rm i}$, is found from the
integral
\begin{equation}
F_{\rm M}(\Delta_{\rm i})=\int_{0}^{\Delta_{\rm i}}d\Delta P_{\rm
V}(\Delta)\Delta,
\end{equation}
where $P_{\rm V}(\Delta)$ is the volume weighted probability
distribution for $\Delta$.  \citet{Miralda2000} quote a fitting
function that provides a good fit to the volume weighted probability
distribution for the baryon density in cosmological hydrodynamical
simulations.  In what follows, we draw primarily from the prescription
of \citet{Miralda2000} and refer the reader to the original paper for
a detailed discussion of its motivations and assumptions.
\citet{Wyithe2003} employed this prescription within a semi-analytic
model of reionization. This model was extended by
\citet{Srbinovsky2007} and by \citet{Wyithe2008}. We refer the reader
to those papers for a full description.

The quantity $Q_{\rm i}$ is defined to be the volume filling factor
within which all matter at densities below $\Delta_{\rm i}$ has been
ionized. The reionization history is quantified by the evolution of
$Q_{\rm i}$ that follows the rate equation
\begin{eqnarray}
\label{preoverlap}
\nonumber \frac{dQ_{\rm i}}{dz} &=& \frac{1}{n_0 F_{\rm M}(\Delta_{\rm
i})}\frac{dn_{\gamma}}{dz}\\
 &-&\left[\alpha_{\rm
B}(1+z)^3R(\Delta_{\rm i})n_0\frac{dt}{dz}+\frac{dF_{\rm M}(\Delta_{\rm i})}{dz}\right]\frac{Q_{\rm
i}}{F_{\rm M}(\Delta_{\rm i})},
\end{eqnarray} 
where $\alpha_{\rm B}$ is the case B recombination coefficient, $n_0$
is the mean comoving density of hydrogen in the IGM, and
$R(\Delta_{\rm i})$ is the effective clumping factor of the IGM. The
evolution is driven by the rate of emission of ionizing photons per
co-moving volume $dn_\gamma/dz$. Within this formalism, the epoch of
overlap is precisely defined as the time when $Q_{\rm i}$ reaches
unity. Prior to the overlap epoch we must solve for both $Q_{\rm i}$
and $F_{\rm M}$ (or equivalently $\Delta_{\rm i}$).  The relative
growth of these depends on the luminosity function and spatial
distribution of the sources. In this regime we assume $\Delta_{\rm i}$
to be constant with redshift before the overlap epoch and compute
results for models with values of $\Delta_{\rm i}\equiv\Delta_{\rm
c}=10$. Different values of $\Delta_{\rm c}$ are not found to
quantitatively effect our results for values in the range
$5<\Delta_{\rm c}<20$ \citep{Wyithe2008}.

Following overlap we may describe the post-overlap evolution of the
IGM by computing the evolution of the ionized mass fraction according
to the equation
\begin{equation}
\label{postoverlap}
\frac{dF_{\rm M}(\Delta_{\rm i})}{dz} =
\frac{1}{n_0}\frac{dn_\gamma}{dz}-\alpha_{\rm
B}(1+z)^3R(\Delta_{\rm i})n_0\frac{dt}{dz}.
\end{equation}
This follows directly from equation~(\ref{preoverlap}) with $Q_{\rm
i}=1$. In this post overlap regime the value of $\Delta_{\rm i}$ is
the dependent variable describing the ionization state of the IGM
(whereas prior to overlap $\Delta_{\rm i}=\Delta_{\rm
c}$). Equation~(\ref{postoverlap}) is integrated to obtain $F_{\rm M}$
(or equivalently $\Delta_{\rm i}$) as a function of redshift.

The emission rate of ionizing photons per co-moving volume that is
required to compute the reionization history can be written
\begin{equation}
\label{star_ionize}
\frac{dn_{\rm \gamma}}{dz} = N_{\rm \gamma}\frac{\langle f_{\rm esc}\dot{\rho}_\star\rangle}{m_{\rm p}}\frac{dt}{dz},
\end{equation}
where $N_{\rm \gamma}$ is the number of ionizing photons produced per
baryon incorporated into stars, and $\langle f_{\rm
esc}\dot{\rho}_\star\rangle$ is the product of the escape fraction and
the star formation rate density averaged over halo mass. As described
in the introduction, only a fraction $f_{\mathrm{esc}}$ of the
ionizing photons produced by stars enter the IGM. We define $\langle \dot{\rho}_{\rm
\star,esc}\rangle$ using our formalism as
\begin{eqnarray}
\nonumber
&&\hspace{-7mm}\langle f_{\rm esc}\dot{\rho}_\star\rangle_{[>M_1]}  = \\
\nonumber
&&\hspace{-7mm}\int_{M_1}^{\infty} dM\,f_{\rm esc} \,SFR\times \frac{\langle t_{\rm ion}\rangle}{t_{\rm H}} \left(t_{\rm H}\Delta M\left.\frac{dN^2_{\rm merge}}{dtd\Delta M}\right|_{M_1,\Delta M}\frac{dn}{dM}\right),
\end{eqnarray}
where $\langle t_{\rm ion}\rangle$ is the time during which the galaxy emits ionizing
photons into the IGM (equation~\ref{tion}). 

In a cold neutral IGM beyond the redshift of reionization, the
collapsed fraction should be computed for halos of sufficient mass to
initiate star formation. The minimum virial temperature is set by the
temperature $T_{{\mathrm{min}}}\sim 10^4\,$K, above which atomic
hydrogen cooling promotes star formation. Following the reionization
of a region, the Jeans mass in the heated IGM limits accretion to
halos above $T_{{\mathrm{ion}}}\sim10^5$K \citep{Efstathiou1992,
Thoul1996, Dijkstra2004b}. Including these separate components from
ionized and neutral regions of IGM, we get
\begin{equation}
\label{SFR_equation}
\langle f_{\rm esc}\dot{\rho}_\star\rangle=\langle f_{\rm esc}\dot{\rho}_\star\rangle_{[>M_{\rm min}]}(1-Q_{\rm i}) + \langle f_{\rm esc}\dot{\rho}_\star\rangle_{[>M_{\rm ion}]}Q_{\rm i},
\end{equation}
\noindent

\begin{figure*}
\begin{center}
\vspace{3mm}
\includegraphics[width=15cm]{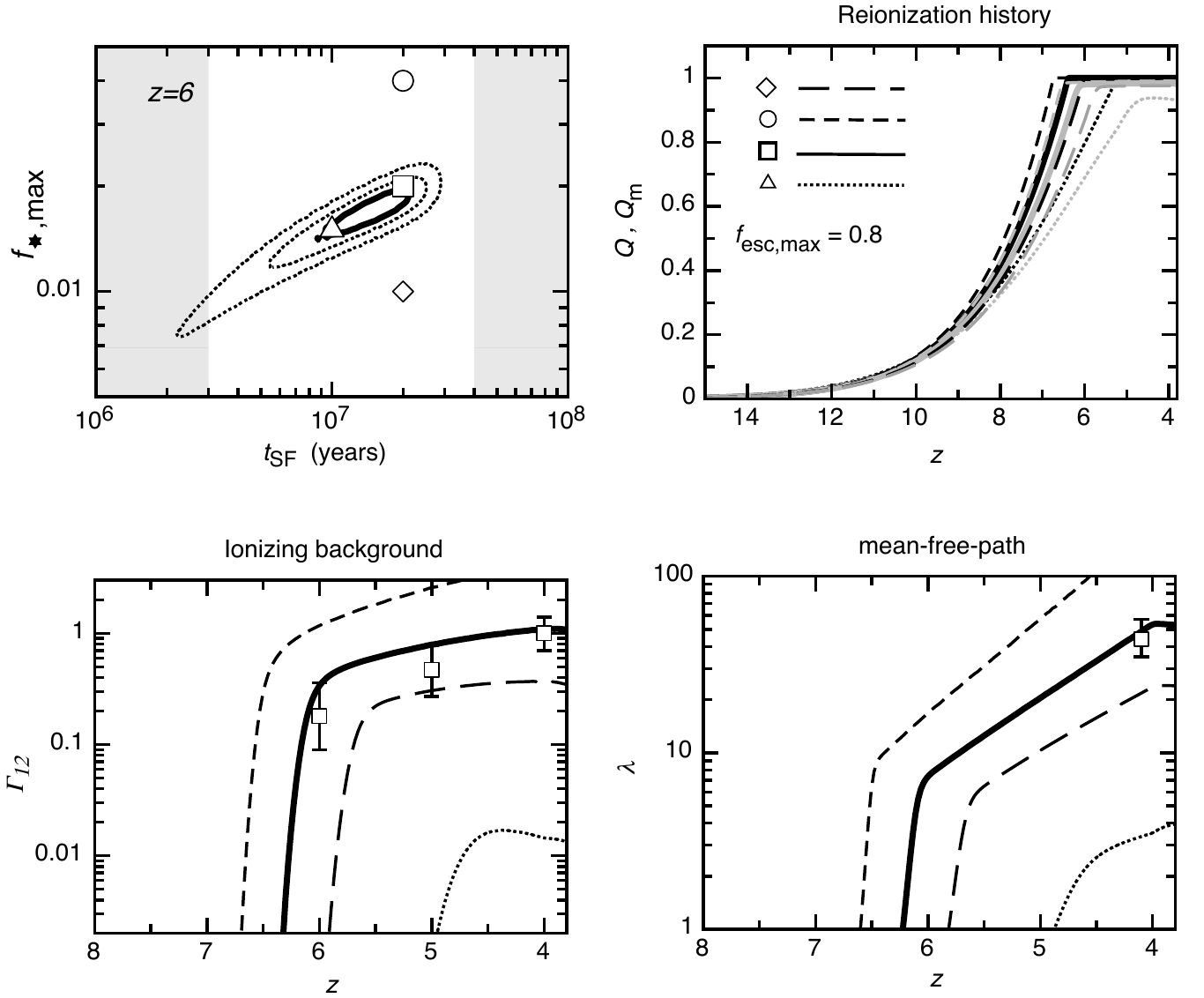}
\caption{\label{fig8} Models for the reionization of the IGM and the
subsequent post-overlap evolution of the ionizing radiation
field. Four cases are shown, with line styles corresponding to the
cases over plotted on the constraints on $f_{\rm \star,max}$
and $t_{\rm SF}$ at $z\sim6$ that are repeated in the top left panel for
reference. A value of $\Delta_{\rm c} = 10$ was adopted for the
critical overdensity prior to the overlap epoch.  {\em Upper Right
Panel:} the evolution of volume averaged (dark lines) and mass
averaged (grey lines) neutral fraction. {\em Lower Left Panel:}
Ionization rate as a function of redshift (in units of
$10^{-12}$s$^{-1}$). The observational points are from
\citet[][]{Bolton2007} and \citet[][]{Wyithe2011}. {\em Lower Right
Panel:} The mean-free-path for ionizing photons. The data point is
based on \citet[][]{Storrie1994}.  }
\end{center}
\end{figure*}
\noindent Our model assumes a spectral energy distribution of Population-II
stars with a metallicity $Z=0.05Z_\odot$ and a \citet[][]{Scalo1998}
initial mass function, for which the resulting number of hydrogen
ionizing photons per baryon incorporated into stars is
$N_\gamma\sim4000$ \citep[][]{Barkana2001}.

In order to estimate the ionizing background following the end of
reionization, we compute a reionization history given a particular
value of $\Delta_{\rm c}$, combined with assumed values for $f_\star$
and $f_{\rm esc}$. Given this history, we then compute the evolution
of the background radiation field due to the same sources.  After the
overlap epoch, ionizing photons will experience attenuation due to
residual over dense pockets of HI gas.  We use the prescription of
\citet{Miralda2000} to estimate the ionizing photon mean-free-path.
Following \citet{Oh2005}, who note that the the constant of
proportionality relating mean-free-path to the volume-filling fraction
should be reduced by approximately a factor of 2, we adopt $\lambda=
(30\mbox{km\,s}^{-1})/H(1-Q)^{-2/3}$, and subsequently derive the
attenuation of ionizing photons. We then compute the flux at the
Lyman-limit in the IGM due to sources immediate to each epoch, in
addition to redshifted contributions from earlier epochs.

\begin{figure*}
\begin{center}
\vspace{3mm}
\includegraphics[width=15cm]{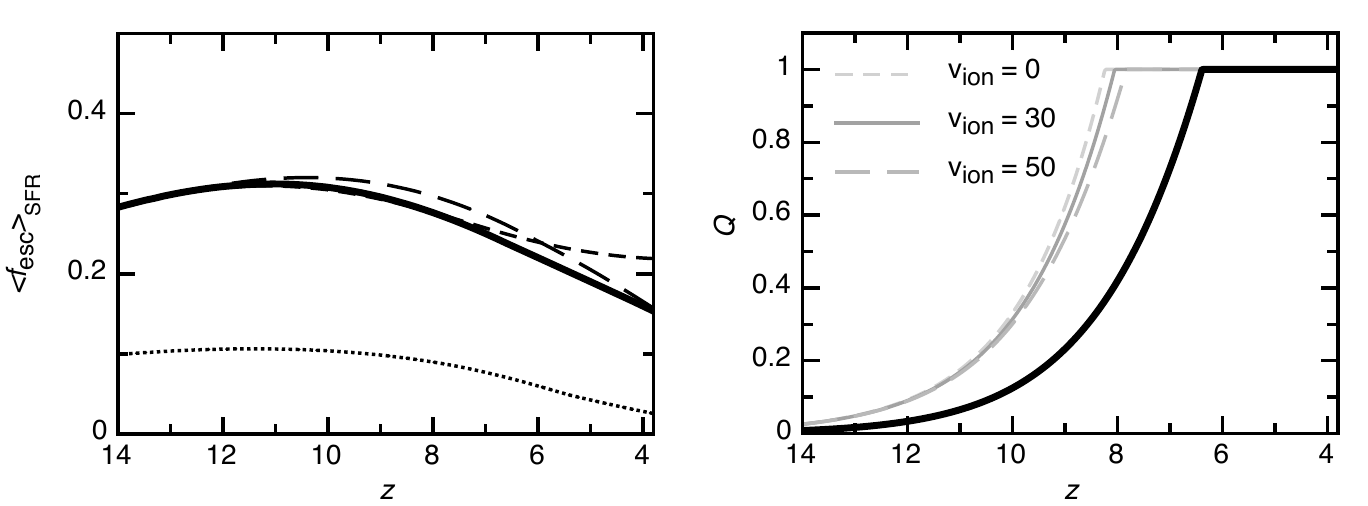}
\caption{\label{fig9} Contribution of galaxies to
reionization.  {\em Left Panel:}    The evolution of $SFR$
weighted average escape fraction for the four models shown in Figure~\ref{fig8}. {\em Right Panel:}  The evolution of volume
averaged neutral fraction. Three cases are shown, including the best
fit model from Figure~\ref{fig8} (thick black line)
for comparison with models in which photons are allowed to escape from
all galaxies, with $f_{\rm esc}=0.8$ (i.e. $SFR_{\rm crit}=0$). In
these cases we show examples where radiative feedback is included
($v_{\rm ion}=30$km/s and $v_{\rm ion}=50$km/s), and a case in which
it is not (i.e.  $v_{\rm ion}=0$).  }
\end{center}
\end{figure*}

We show the results of this modelling in Figure~\ref{fig8} for four
models, including the best-fit case presented in Figure~\ref{fig1}.
For reference the constraints on $f_{\rm \star,max}$ and $t_{\rm SF}$
at $z\sim6$ are repeated in the top left panel. The upper right panel
shows the evolution of volume averaged (dark lines) and mass averaged
(grey lines) neutral hydrogen fraction. To illustrate that the model for star
formation is consistent with the reionization history, we also show the
observables of ionization rate as a function of redshift (lower left), and the
mean-free-path for ionizing photons (lower right). We find for a choice
of $f_{\rm esc}=0.8$, that a model near the best fit for the SFRD
function ($t_{\rm SF}=2\times10^7$yr) results in a reionization
history that is consistent with low redshift constraints (thick
lines). The reionization history shown is consistent with recent constraints from the patchy kinetic Sunyaev-Zel'dovich effect \citep[][]{Zahn2012}.  The model produces an optical depth to electron
scattering of $\tau_{\rm es}=0.065$, which is lower than the observed value of $\tau_{\rm es}=0.088\pm0.015$
\citep[][]{Komatsu2011}.  However, our model does not include the
possibilities of population-III stars or an increased escape fractions
at very high redshift~\citep[][]{Alvarez2012}. Models with larger $f_{\rm \star,max}$
reionize the universe too early, while those with smaller $f_{\rm \star,max}$
or larger $t_{\rm SF}$ reionize the universe too late. For the model with a smaller value of $t_{\rm SF}=10^7$yr, we
find that reionization occurs very late (at $z\sim5$), and greatly
underestimates the ionizing background. This is because porosity of
unity is rarely achieved by the time the starburst is finished in
this model, so that very few ionizing photons escape. This indicates
that the reionization of the IGM may limit the starburst time-scale
to be $t_{\rm SF}\gtrsim10^7$yr.

In the left panel of Figure~\ref{fig9} we show the $SFR$
weighted average of the escape fraction (i.e. $\langle f_{\rm
esc}\rangle_{SFR}\equiv \langle f_{\rm esc}\dot{\rho}_\star\rangle/
\langle \dot{\rho}_\star\rangle$) as a function of redshift for the four models shown in Figure~\ref{fig8}. Since the
value of escape fraction in highly star forming galaxies is $f_{\rm
esc}=0.8$ in this model, we find that most ionizing photons produced
in galaxies are being lost either because the $SFR$ was insufficient
to attain porosity equal to unity, or because the UV luminous stars
had died before porosity of unity was achieved. Interestingly, we find
that the SFR averaged escape fraction increases towards high redshift,
being twice as large at $z\sim10$ as at $z\sim4$. This trend is
consistent both with the requirements of the Ly$\alpha$ forest at
$z\sim4$ \citep[][]{Bolton2007}, and recent empirical estimates for
the escape fraction of ionizing photons from high redshift galaxies
during reionization \citep[][]{Finkelstein2012}.

\begin{figure*}
\begin{center}
\vspace{3mm}
\includegraphics[width=15cm]{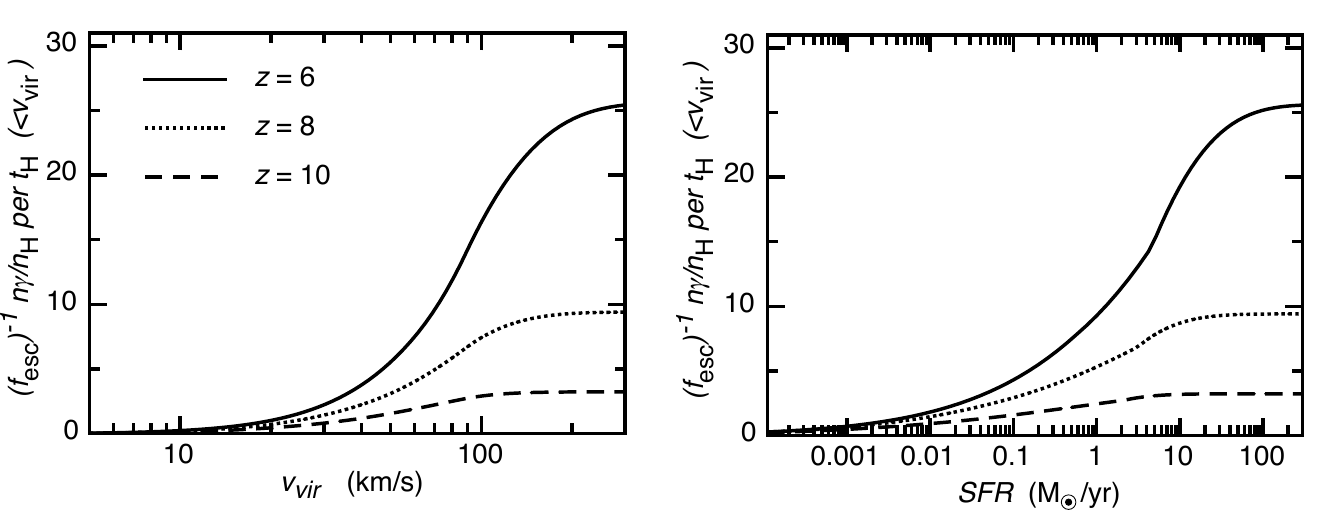}
\caption{\label{fig10} Contribution of low mass galaxies to
reionization.  {\em Left Panel:} The cumulative contribution to the number of ionizing
photons per hydrogen in the Universe as a function of halo virial
velocity. To investigate the possible contribution to reionization
from small galaxies we have not applied the $SFR_{\rm crit}$ criteria,
and instead assumed that ionizing photons escape all galaxies with
$f_{\rm esc}=0.8$. Contributions at three different redshifts are
shown, and expressed in units of photons per hydrogen in the IGM times
$f_{\rm esc}^{-1}$ (in order to make the curves independent of $f_{\rm
esc}$). {\em Right Panel:} Results repeated as a function of the
$SFR$ per galaxy. }
\end{center}
\end{figure*}

\section{The contribution of low mass galaxies to reionization}
\label{contribution}

A range of models \citep[e.g.][]{Haiman2003,Iliev2007} have predicted that
radiative feedback, which can suppress star formation in very low mass
galaxies, will have a significant effect on the reionization
history. The effect is often referred to as self regulation of
reionization, because it could lead to the reionization process being
extended. In these models reionization is predicted to start at high
redshift due to very low mass galaxies with virial temperatures of
$\sim10^4$K. However, once regions become ionized subsequent star
formation in these very low mass galaxies is quenched, so that
reionization must be completed by galaxies with a virial temperature
$\gtrsim10^5$K which form later. The reason why radiative feedback
could have a large effect on the reionization history is well
understood. At high redshift the collapsed fraction of dark matter
halos is dominated by low mass galaxies with a virial temperatures
smaller than $\sim10^5$K. Despite the focus of numerical and
semi-analytic studies of high redshift galaxy formation on the role of
feedback on star formation in low mass galaxies
\citep[e.g.][]{Raicevic2011,Finlator2011}, models of the structure of
reionization have generally assumed the luminosity to be proportional
to halo mass, and that the escape fraction of ionizing photons is
independent of mass \citep[e.g.][]{Iliev2007,McQuinn2007}.  As a
result, the ionizing photon budget for reionization in these constant
mass-to-light models is also dominated by low mass galaxies prior to
reionization. However, in this paper we have found that SNe feedback is
required to reproduce the shape of the high redshift SFRD function, in
agreement with the previous numerical work. Thus, the contribution to
reionization from low mass galaxies may be smaller than previously
thought \citep[][]{Raicevic2011}. In this section, we investigate the
contribution to reionization from galaxies of various masses.

The effect of low mass galaxies on the reionization history is
illustrated in the right panel of Figure~\ref{fig9} which shows
the evolution of volume averaged neutral fraction. The best fit model
from Figure~\ref{fig8} is shown (thick black line) for comparison
with models in which the critical SFR criteria for ionizing photons to
escape is not applied so that photons are allowed to escape from all
galaxies with $f_{\rm esc}=0.8$ (grey lines). For the latter, three
cases are shown including where radiative feedback is not included
(i.e. $v_{\rm ion}=0$) and where $v_{\rm ion}=30$km/s and $v_{\rm
ion}=50$km/s. We see that the application of the critical SFR criteria
has a significant effect on the reionization history, leading to an
earlier reionization by $\Delta z\sim1.5$ at a fixed $f_{\rm
esc}$. However, suppression of galaxies smaller than $v_{\rm
ion}=30$km/s or 50km/s results in a smaller change to the reionization
history relative to a model in which radiative feedback does not
operate ($\Delta z\lsim0.2$), a finding that is contrary to the
standard picture for self-regulated reionization
\citep[e.g.][]{Iliev2007}. 

To understand the reasons for this we show the cumulative contribution
to the number of ionizing photons per hydrogen in the Universe per
Hubble time as a function of halo virial velocity (left panel of
Figure~\ref{fig10}). In calculating these curves we have not applied
the $SFR_{\rm crit}$ criteria (equation~\ref{sfrcrit}), and assumed
that ionizing photons escape all galaxies with $f_{\rm
esc}$. Contributions at three different redshifts are shown. In the
right panel we repeat these results, but plot the contributions
as a function of star formation rate rather than virial velocity. We
find that photons from galaxies below $v_{\rm vir}\sim30$km/s (with
$SFR\la0.01$M$_\odot$/yr), corresponding to those affected by
radiative feedback, represent only $\sim10\%$ of the potential photon
budget at $z\sim6$ and only $\sim30\%$ at $z\sim10$. Thus, these low
mass galaxies make only a small contribution to reionization, implying
that radiative feedback should not be important in regulating the
reionization process. This reduced contribution arises because SNe
have lowered the star formation efficiency to a level where galaxies
with $v_{\rm vir}\sim10$km/s contribute a very small fraction of the
total star formation rate, despite the corresponding halos
representing the dominant component of collapsed fraction of dark
matter.

Thus, we conclude that SNe feedback was much more important than
radiative feedback in shaping the reionization history, a result that
is consistent with both \citet[][]{Raicevic2011} and the recent
semi-analytic modelling of \citet[][]{Kim2012} using GALFORM within
the Millennium-II simulation. We note that an important caveat to
these results is the possibility of a top-heavy mass function of
Population-III stars in small galaxies. Since in this case the
ionizing efficiency is much larger than for Population-II stars,
Population-III stars in early low-mass galaxies could make a more
important contribution to reionization, partially reionizing the
Universe at earlier times \cite[e.g.][]{Wyithe2003,Cen2003}.
Overall we find that half of the ionizing photons are produced by
galaxies with $v_{\rm vir}\gtrsim 60$km/s corresponding to $SFR\gtrsim
1$M$_\odot$/yr. These are observed galaxies, indicating that
observations summarised in \citet[][]{Smit2012} correspond to about
half of the ionizing photons produced by galaxies. This result is
consistent with the recent findings of \citet[][]{Finkelstein2012},
who suggested that the observed population of $z\sim6$ galaxies is
sufficient to provide most of the ionizing photons required for
reionization assuming escape factions below $50\%$.

\section{Conclusion}
\label{conclusion}

We have described a simple model for the star formation rate density
function at high redshifts based on the extended Press-Schechter
formalism. This model assumes that a starburst is triggered by each
major merger, lasting for a time $t_{\rm SF}$ and converting at most
$f_{\rm \star,max}$ of galactic gas into stars. We include a simple
physical prescription for SNe feedback based on the galactic porosity
model of \citet[][]{Clarke2002} that suppresses star formation in low
mass galaxies, and results in a minimum galaxy mass from which
ionizing photons can escape. This model for star formation has only
two free parameters, $t_{\rm SF}$ and $f_{\rm \star,max}$, but
accurately describes recent measurements of the amplitude and shape of
the star formation rate density function between redshifts 4 and 7.

Comparison of our modelling with observational data implies that
individual starbursts were terminated after $t_{\rm
SF}\sim2\times10^7$ years. This termination time lies between the
main-sequence lifetimes of the lowest and highest mass SNe
progenitors, indicating that starbursts were quenched once SNe
feedback had time to develop. From the sum of all major merger events,
high redshift galaxies convert $\sim5-10\%$ of their mass into stars for
large galaxies with star formation rates above
$\sim1$M$_\odot$/yr. However, the overall star formation efficiency of
lower luminosity galaxies is only a few percent. In our model, high
redshift galaxies have a high duty-cycle for star formation,
undergoing starbursts $\sim10\%$ of the time at $z\sim 10$.

We calculate the relation between stellar mass and star formation
rate, finding it to be approximately linear, in agreement with
previous theoretical work and observations. The predicted ratio of
star formation rate to stellar mass based on our fit to the star
formation rate density function agrees with the observed values for
star-forming galaxies at $z\sim4$--$6$ where stellar mass has been
measured. Moreover, if the Milky-Way dwarf spheroidals represent
fossil records of the star-forming galaxies during reionization
\citep[][]{Rocha2012}, they should also have stellar masses described
by our model and we find good agreement with the predicted relation
between stellar and halo masses (although the scatter is large). This
implies that our model correctly describes the stellar mass to halo
mass ratio from $10^5$M$_\odot\lsim M_\star\lsim10^{10}$M$_\odot$.

We used our model to discuss the escape fraction of ionizing photons
from high redshift star-forming galaxies.  \citet[][]{Clarke2002}
proposed a simple model where galaxies with a sufficient star
formation rate to generate a porosity greater than unity had an escape
fraction for ionizing photons that is of order unity, whereas galaxies
with insufficient start formation have a negligible escape
fraction. Since a fraction of the galaxies starburst lifetime is
required to build the porosity to unity, every star-forming galaxy has
a non-zero probability of being observed with a zero escape fraction
for ionizing radiation. Our model predicts that large escape fractions
should be rare for both low SFR and high SFR galaxies. However, we
expect approximately a half of star-forming galaxies with
$SFR\sim0.1-1$M$_\odot$/yr to have significant escape fraction. Thus
our model provides a natural explanation for the wide range of
conclusions regarding observations of the escape fraction of ionizing
photons from star-forming galaxies \citep[e.g.]{Steidel2001,
Fernandez2003, Shapley2006, Siana2007}, and predicts that the escape
fraction during reionization at $z\sim10$ was twice as large as at
$z\sim4$.
 
We also used a semi-analytic model for the reionization process based
on our SNe regulated star formation history. Even after allowing for
the fact that SNe feedback enables escape of ionizing radiation only
after the galactic porosity reaches unity, we find that our model is
able to reionize the Universe within current observational
constraints.  We find that low mass galaxies were minor contributors
to reionization, owing to the suppression of star formation by SNe
feedback. We further find that this SNe feedback lowers the efficiency
of star formation in low mass galaxies to such an extent that
photo-ionization feedback on low mass galaxy formation does not
significantly effect the reionization history. This is because the
galaxies that would have been subject to radiative feedback are only
very minor contributors to the potential ionizing photon budget once
SNe feedback is taken into account. \footnote{An important caveat to
this conclusion is the possibility of a top-heavy mass function of
Population-III stars in small galaxies, since in this case the
ionizing efficiency is much larger than for Population-II stars, so
that low-mass galaxies could make a more important contribution to
reionization in this case.} Finally, we find that approximately half
of the ionizing photons needed to complete reionization have already
been observed in star-forming galaxies at $z=6$--10.

\vspace{5mm}

{\bf Acknowledgments} We thank Jamie Bolton for helpful discussions.
JSBW acknowledges the support of the Australian Research Council.  AL
was supported in part by NSF grant AST-0907890 and NASA grants
NNX08AL43G and NNA09DB30A.

\newcommand{\noopsort}[1]{}

\bibliographystyle{mn2e}

\bibliography{text}

\label{lastpage}
\end{document}